# In-plane orientation effects on the electronic structure, stability and Raman scattering of monolayer graphene on Ir(111)

Elena Starodub,[1] Aaron Bostwick,[2] Luca Moreschini,[2] Shu Nie,[1] Farid El Gabaly,[1] Kevin F. McCarty[1] and Eli Rotenberg[2]

[1]Sandia National Laboratories, Livermore, California 94550, USA
[2]Advanced Light Source, E. O. Lawrence Berkeley National Laboratory, Berkeley, California 94720, USA

We employ angle-resolved photoemission spectroscopy (ARPES) to investigate the electronic structures of two rotational variants of epitaxial, single-layer graphene on Ir(111). As grown, the more-abundant R0 variant is nearly charge-neutral, with strong hybridization between graphene and Ir bands near the Fermi level. The graphene Fermi surface and its replicas exactly coincide with Van Hove singularities in the Ir Fermi surface. Sublattice symmetry breaking introduces a small gap-inducing potential at the Dirac crossing, which is revealed by *n*-doping the graphene using K atoms. The energy gaps between main and replica bands (originating from the moiré interference pattern between graphene and Ir lattices) is shown to be non-uniform along the mini-zone boundary due to hybridization with Ir bands. An electronically mediated interaction is proposed to account for the stability of the R0 variant. The variant rotated 30° in-plane, R30, is *p*-doped as grown and K doping reveals no band gap at the Dirac crossing. No replica bands are found in ARPES measurements. Raman spectra from the R30 variant exhibit the characteristic phonon modes of graphene, while R0 spectra are featureless. These results show that the film/substrate interaction changes from chemisorption (R0) to physisorption (R30) with in-plane orientation. Finally, graphene-covered Ir has a work function lower than the clean substrate but higher than graphite.

1. **Introduction**

Graphene, a single-layer sheet of $sp^2$-bonded graphitic carbon, is a wonderful material to study the basic physics of a two-dimensional semimetal. Its massless charge carriers, Dirac fermions, lead to fascinating transport properties.[1] Characteristics such as high carrier mobility,[2] a width-dependent band gap in nanoribbons[3] and the ability to manipulate the band gap in bilayer graphene with chemical doping[4] or external electric field[5] offer much potential for next-generation electronics.[6]

Free-standing graphene has an electronic structure in which π and π* bands touch at the Fermi energy ($E_F$) at the K and K' corners of the Brillouin zone. Near these Dirac points, the bands exhibit linear dispersion.[7, 8] However, intimate contact with a substrate changes graphene's electronic structure. Many experimental studies on the electronic structure of supported graphene have used SiC(0001) substrates.[9-11] Graphene on metals is of interest as a route to synthesizing high-quality graphene[12, 13] and for electrical contacts to devices.[14] Graphene/metal systems can be roughly classified into two different categories of film/substrate binding — weak (physisorbed) or strong (chemisorbed).[13] On some metals with strong binding, like Ni(111)[15] and Ru(0001),[16] graphene grows only with one in-plane orientation. On other metals with weaker binding, like Pt(111),[17] Pd(111)[18] and Ir(111),[19] graphene has several azimuthal orientations. How in-plane orientation affects properties is not well-known. The orientation on Pt(111) did not measurably affect graphene's electronic structure.[20] In contrast, orientation changes the work function of graphene-covered Pd(111),[18] a more strongly interacting system.[14] Here we focus on understanding how orientation affects electronic structure and Raman scattering by studying graphene on Ir(111). In this system,[12, 21-23] graphene has at least six different in-plane orientations.[19, 24, 25]

The electronic structure of graphene on Ni(111),[26-28] Ru(0001)[29-32] and Ir(111)[33-35] has been studied in detail for the case where the graphene and substrate lattice vectors are aligned in-plane. We label this crystallographic orientation as R0 graphene. For single-layer graphene on the strongly interacting metals Ni and Ru, the π band is lowered in energy and a band gap opens at the K and K' points.[13, 26] Localized states are observed in the gap.[13] In contrast, aligned (R0) graphene on Ir(111) has almost the electronic structure of free-standing graphene, as reported by Pletikosić et al.[33] However, a superperiodic potential resulting from the mismatch of the graphene and Ir lattices was found to affect the electronic structure by creating replicas of the Dirac cone and opening "mini-gaps" between the primary and replica Dirac cones. These mini-gaps were reported to have a strong variation along the reduced Brillouin zone boundary by Pletikosić et al.,[33] who were not able to resolve if a band gap also existed at the Dirac point $E_D$. However, they concluded that any gap at $E_D$ would have to be less than 200 meV. Although Balog et al. showed that hydrogen induced a gap,[35] they reported no gap for clean graphene on Ir(111).

Here we explore R0 graphene on Ir in more detail by varying the photon energy in angle-resolved photoemission spectroscopy (ARPES) and intentionally doping the material. We find that the π band of the R0 variant is strongly hybridized with an Ir 5$d$ state near the Fermi level, creating a gap between the π and π* bands. Doping with K removes the hybridization near the Dirac point, revealing the residual band gap due to sublattice symmetry breaking.

We then investigate the effect of in-plane orientation by studying the rotational variant that is rotated 30º in-plane, R30 (see schematic in Fig. 1). Compared to R0, the less-stable R30 variant[36] has weaker hybridization with Ir bands near $E_F$, has no band gap, is *p*-doped by the substrate and exhibits no replicas of the Dirac cone and thus no mini-gaps. We observe the graphene vibrational modes in Raman spectra from the R30 variant, unlike the R0 variant. These results show that R0 graphene is more strongly bound to Ir than R30 graphene and highlights the sensitivity of graphene's properties to in-plane orientation on a metal. Finally, covering Ir with graphene lowers the work function, but not below the value of graphite.

2. **Experimental methods**

Two types of monolayer graphene films were prepared. The first was a mixture of R30 and R0 domains; the second contained only R0 graphene. These films were grown on Ir(111) by a combination of decomposing ethylene and segregating carbon from the Ir during cooling.[19, 37] By directly observing film growth using low-energy electron microscopy (LEEM), we optimized the growth conditions to produce a very low density of nuclei separated by several tens of microns.[19] These nuclei were then expanded to cover the surface while minimizing secondary nucleation. The separately nucleated domains had a high degree of in-plane registry, as characterized *in situ* using selected-area, low-energy electron diffraction (LEED).[19] Since R30 islands are harder to nucleate but grow faster,[37] phase-pure R0 films were produced by growing very slowly. Fast growth led to a significant fractional coverage by R30 graphene. Work functions were determined at room temperature by collecting a series of LEEM images as a function of the incident electron energy.

For ARPES, graphene-covered Ir(111) samples were transferred in air from the LEEM apparatus to the Electronic Structure Factory endstation on beamline 7.0.1 at the Advanced Light Source[4] and annealed in ultrahigh vacuum to remove adsorbates from the air exposure. The ARPES measurements were conducted with the sample at ~20 K using 95 and 130 eV photons with an overall energy resolution of ~25 meV. The as-grown graphene was doped at ~20 K by depositing potassium from a commercial SAES getter source. After doping, the replicas of the R0 K points were still present, establishing that the potassium was adsorbed and not intercalated.

Raman spectra were acquired *ex situ* in a 180°-backscattering geometry using a 100× objective lens and 532-nm excitation from a frequency-doubled Nd:YAG laser. The scattered light was filtered by a film polarizer in crossed polarization with the incident, linearly polarized light to reduce the background scattering.[38] A Semrock edge filter was used to reject the elastically scattered light. A spectrograph with a single 600 groove/mm grating dispersed the light onto a CCD detector cooled by liquid nitrogen. The laser spot size was approximately ~1 μm in diameter. The spectrometer was calibrated with a neon lamp. Raman maps were collected by scanning the sample stage.

3. **Results and discussion**

3.1. Dirac points of R0 and R30 graphene

At a binding energy slightly below $E_F$, Fig. 2a maps in two momentum directions the constant-energy contours observed by ARPES from a film containing a mixture of R0 and R30 graphene. $K_{R0}$ and $K'_{R0}$ label the R0 K points. Figure 1a illustrates the Brillouin zones (BZs) of the R0 variant and Ir(111), following Pletikosić et al.[33] Besides the primary bands of single-layer R0 graphene, we identify six replicas of the R0 π states surrounding each $K_{R0}$ and $K'_{R0}$ point. As described by Pletikosić et al.,[33] these replicas come from the periodic potential created by the mismatch of the R0 graphene and Ir lattices. Since 10 carbon atoms lie over about 9 Ir atoms,[39] the film plus substrate create a moiré (coincidence) lattice with reciprocal lattice vectors $\mathbf{c}^{moiré}$, which are the difference between the Ir and graphene reciprocal lattice vectors, $\mathbf{a}^{Ir}$ and $\mathbf{b}^{R0}$. The (10 × 10) superstructure spots in LEED patterns[21, 39] also result from the moiré superstructure.

The intensity distribution of each graphene Fermi surface contour is non-uniform as a consequence of the interference of electrons emitted from the two sublattices of graphene.[40-42] Thus, the individual Fermi surfaces appear as horse-shoe shaped arcs. Note that the horseshoe-shaped arcs of the replicas do not point in the same direction as the primary arcs, unlike monolayer graphene on SiC.[9] In SiC, the replica bands are due to final-state diffraction[41] but in the present case, the rotation of the arcs suggests the introduction of a quantum-mechanical phase term in the initial state. A similar, but smaller rotation was observed for 2 ML of graphene on Ru(0001).[32] In contrast to the intensity distributions, the trigonal warping of the replica bands (the deviation of the Fermi surface contours at large energy scales due to the lattice) is aligned with that of primary bands.

Additional, nearly circular features in the band structure, marked by $K_{R30}$ and $K'_{R30}$ in Fig. 2, are detected rotated 30° from $K_{R0}$ and $K'_{R0}$. These states, which are not observed in similar data from a pure R0 film, are the $\pi$ bands arising from minority domains of R30 graphene. No replicas are observed around these R30 K points, in contrast to the R0 variant. The black open circles in Fig. 1b show the replicas around one $K_{R30}$ point expected from the R30 superstructure on Ir(111). Assuming commensurability, the $\mathbf{c}^{\text{moiré}}$ vectors of the ($\sqrt{124}\times\sqrt{124}$)-R9° moiré cell[19] are rotated by 9° from the close-packed Ir directions.

Since the Dirac points of the R0 and R30 variants are well-separated in reciprocal space, we were able to map the spatial distribution of variants on the surface using the photoemission intensities at a K point of each variant. In this way we found regions of each variant larger than the x-ray beam and obtained phase-pure spectra from our mixed-variant film.

3.2. Energy bands of R0 graphene

For as-grown R0 graphene, Fig. 3(a, left) shows the energy bands and Fermi surface around a K point for photon energy 95 eV. The intense central linear feature in the left panels is the graphene π band measured along the $\Gamma K_{R0}$ direction, whose sharpness shows that the graphene has high crystalline quality. The presence of a single π band in the left panel reveals that the graphene consists of a single layer,[43] consistent with our *in situ* LEEM and LEED characterization. Figure 3(a, center) shows the energy bands along the perpendicular momentum direction through the graphene K point. In this geometry, both segments of the slice through the cone-shaped π are revealed by photoemission.[41, 42] Linearly extrapolating the bands from higher binding energy reveals a projected Dirac crossing ~100 meV above $E_F$, consistent with Pletikosić et al.[33] This group suggested the possibility of a band gap in R0 graphene. However, as the data in Fig. 3(a, left) show, the graphene bands in fact are strongly distorted near $E_F$, so that an extraordinary kink in the energy band can be observed with a clear metallic Fermi crossing. This is reflected in the Fermi surface shown in Fig. 3(a, right), which is shifted far to the right from the graphene K point and distorted from the point-like feature expected for non-interacting graphene.

This distortion results from an accidental degeneracy of the graphene and Ir Fermi bands near $E_F$, as shown in Fig. 2b-d. Figure 2b shows a magnified view of the constant-energy contours 150 meV below $E_F$. Ir 5*d* states in this region have a three-fold shaped energy gap, indicated by the dashed lines, centered on the Ir lattice K point.[44] In the reduced symmetry of the graphene-Ir system, the graphene Dirac cones exist at the center (Γ point) of a "mini-Brillouin Zone" (mini-BZ).[33] As the binding energy approaches $E_F$, the gap in the Ir states is pinched off as the binding energy approaches $E_F$, resulting in a saddle point of Ir states accidentally degenerate with the Γ point of the mini-BZ. This results in a strong hybridization of the graphene and Ir states, removing spectral weight from the graphene K point (Γ point of the mini-BZ), as observed.

To support this picture, we now examine how the energy bands and Fermi surface change with photon energy and doping. The intensity of the π band excited by 130 eV photons shown in Fig. 3(b, left) decreases rapidly near the Fermi level. A comparison with Fig. 3(a, left) shows that the strong, weakly dispersing feature marked by the dashed line near $E_F$ is still present, but greatly suppressed in intensity. In a study of clean Ir(111) using different photon energies and density functional theory calculations (DFT), Pletikosić et al.[44] identified this band as a *d*-like surface state of Ir. These states, highlighted in Fig. 3(a, right) by the dashed lines, form the saddle point nearly exactly at the graphene K point, as mentioned above (Fig. 2b-d). In contrast to the Iridium 5*d* state's pronounced decay of photoemission cross section with photon energy,[45] graphene's energy bands change in intensity only weakly and monotonically with photon energy.[43]

In contrast to the 130 eV excitation (Fig. 3b), the π band excited by 95 eV photons does not diminish in intensity approaching $E_F$ (Fig. 3a). Thus, the π band intensity at $E_F$ is strong (suppressed) when photoemission from the Ir 5*d* band is strong (weak). This intensity scaling of the π and Ir bands shows that they are strongly hybridized. As Fig. 3(a, center) reveals, the as-grown R0 graphene has an arc of intensity that bridges the two π bands near the Fermi level. Thus, near $E_F$ the lower Dirac cone is closed and separated from the upper Dirac cone, which lies above $E_F$. Therefore, as-grown R0 graphene has a band gap that results from hybridization with the Ir band. This gap could be regarded as an anti-crossing between graphene π and Ir 5*d* bands.

Doping the graphene using electrons donated from deposited K atoms reinforces this interpretation. The lower three sets of panels in Fig. 3 show the affect of increasing K dose, all measured using 130 eV photons. The left-hand panels show that the π band intensity extends closer to $E_F$ with increasing K dose. The center panels establish that for sufficiently high doping, the π bands clearly cross $E_F$. These features correspond to the emergence of a clear, ~circular Fermi surface seen in the lower two right panels. This Fermi surface emerges when the π bands are far enough away from the saddle point in the Ir band structure (Fig. 2b-d).

At the Dirac crossing energies $E_D$ in Figs. 3(c-e, left), the π band shows a reduced intensity. This effect could result from a very small (~100 meV) band gap, with differing origin from the hybridization-induced gap discussed above. As pointed out, the most likely explanation for this is sublattice symmetry breaking, which arises naturally in a ~$3m \times 3m$ reduced Brillouin zone of the combined lattices ($m$ is an integer), and which couples K and K' points of graphene.[41][46] Although the approximate 10 × 10 symmetry would not predict such a gap, there is the possibility of discommensuration, as indicated by STM measurements.[39]

3.4 Mini-gap spectrum of R0 graphene

We have also investigated the interaction between main and satellite bands in more detail. This interaction occurs at large binding energies away from $E_F$ and is reflected in the formation of "mini-gaps" along the mini-BZ boundary (Fig. 2).[33] An example of such a mini-gap occurs at a binding energy around -770 meV in Fig. 3(a, center). Previously, a large variation of this gap (by a factor of 2.4) along the BZ boundary was observed,[33] which was suggested to be related to the moiré superlattice potential. Such a variation as previously reported violated the rotational symmetry of the Dirac cones, however. Therefore, we performed an ARPES investigation of the spectrum along the mini-BZ boundary, and the results are shown in Fig. 4.

Figure 4a shows a magnified view of the momentum distribution slice near the mini-BZ at -770 meV binding energy. (In the reduced symmetry of the mini-BZ, the graphene cones occur at Γ points.) The binding energy of Fig. 4a was chosen such that the faint circular contours of the satellite cones overlap at the mini-BZ boundary. We could by analysis of this map determine the mini-BZ boundary quite precisely, within (we estimate) a few percent of the mini-BZ dimensions. The ARPES spectrum along the mini-BZ boundary could be extracted from the full energy-momentum data set and is shown in Fig. 4b. The mini-gaps were identified following Pletikosić et al.[33] and fitted to determine the binding energies of the lower and upper rims, as indicated by the solid and dashed lines, respectively. The fits were carried out only along the points where two clear bands could be observed (along some places one or the other band is very weak, as discussed above).

We observe first of all that the symmetry of the bands is reduced to three-fold. This is due to the trigonal warping of the Fermi surface, which puckers the Fermi surface towards the $K_0$, $K_2$ and $K_4$ points of the mini-BZ (see Fig. 4a). At the top of Fig. 4b, we plot the gap size (the difference between upper and lower band energies) and compare to results of Pletikosić et al. We find that along the lines $K_2$-$K_3$'-$K_4$ a significantly smaller variation in mini-band gap size than seen previously,[33] and in fact could fit the mini-gap size to a constant ~160 meV.

Examining data along the lines $K_2$-$K_1$' and $K_4$-$K_5$', we see a systematic increase in the mini-gap size up to around 230 meV. This is not as large as the variation seen by Pletikosić et al. Also, the trend is reversed from Pletikosić et al., who saw the largest gap at $K_3$', whereas we see the smallest gap there.[47] Still the mini-gap variation points to a reduction in symmetry below the expected three-fold warping — the mini-gap appears smaller at the left side of the mini-BZ and increases away from it — but the symmetry is consistent with the overall symmetry of the graphene-Ir system. We explain the increase in mini-gap size qualitatively by the hybridization of the graphene states to the Ir states on the right side of the mini-BZ. (These Ir states dominate the spectrum between $K_0$-$K_1$' and $K_5$'-$K_0$ in Fig. 4b.)

3.5 Energy bands of R30 graphene

Figure 5 shows graphene band structure data around the R30 K point. Before potassium doping, the π bands in Figs. 5(a, left and center) are nearly straight, symmetric and extend to the Fermi level. A clear Fermi surface is apparent in Fig. 5(a, right), with circular shape expected for doped graphene, in contrast to the situation for the R0 phase, Fig. 3(b, right). The R30 Dirac cones exist near the Ir surface Brillouin zone M point (Fig. 1b); unlike the R0 phase in Fig. 2b-d, no part of the Dirac cones are in a gap in the Ir states projected onto the surface.[44] The extrapolated bands cross at about 175 meV above $E_F$, showing that the R30 graphene is doped p-type. The weaker bands in Fig. 5a are from Ir and they do not interact (hybridize) with the graphene π band to nearly the extent as the R0 phase. The residual hybridization is detected in the slight transfer of intensity from π bands to Ir bands near their crossings at $E_F$ (Fig. 5, right panels), and by an apparent non-linear band dispersion near such crossings in the left and center panels. Small gaps at these crossings may be associated with this hybridization, but such gaps if they exist affect only parts of the graphene Fermi surface, not removing it entirely as in the case of the R0 variant.

Figure 5b shows the R30 variant after doping with K to the same level as Fig. 3e. It cannot be clearly determined whether a sublattice-symmetry induced band gap between the π and π* orbitals exists in the doped R30 variant.

3.6 Raman scattering from R0 and R30 graphene

We next show that the differences in electronic structure of R0 and R30 graphene also affect Raman scattering. The LEEM image in Fig. 6a shows the Ir substrate covered with single-layer R0 and R30 graphene near two crossed scratches used as fiducials for Raman spectroscopy. The regions of bright, dark and medium contrast are R30, bare Ir and R0, respectively. The spectra in Fig. 6b are displaced from bottom to top according to the position along the red arrow shown in Fig. 6a. The bottom spectrum, from the R30 graphene at the start of the arrow, exhibits the characteristic G and G' (2D) peaks of graphene.[48] The next four spectra, from bare Ir or R0 graphene, are featureless. The top four spectra, from the R30 graphene above the horizontal scratch in Fig. 6a, again have the G and G' peaks. Clearly the Raman-active phonons of graphene are only observed from the R30 variant but not the R0 variant.

Background-subtracted spectra are shown in Fig. 6c. The G peak of R30 is at 1598 cm$^{-1}$ with a mean full width at half maximum (FWHM) of 13 cm$^{-1}$. The G peak position is close to the 1597 cm$^{-1}$ value reported for single-layer graphene on SiC.[49] Based on results from electrically doped, non-epitaxial graphene,[50] the measured doping level of as-grown R30 graphene (about a 175 meV shift of the Dirac point, Fig. 5a) would account only for roughly one third of the G peak shift. Therefore, we suggest that the majority of the G peak's blue shift comes from residual compressive stress, despite the fact that stressing-relieving ridges form in graphene on Ir during cooling.[19, 24] A disorder-induced D peak is detected at 1339 cm$^{-1}$ with four times smaller intensity than the G peak, indicating a reasonable film quality. We also find the G' (2D) peak at 2688 cm$^{-1}$, which results from a two-phonon, second-order resonance process.[48, 51] The right panel of Fig. 6a shows the G' peak fitted by a single Lorentzian with FWHM of 45 cm$^{-1}$.

The graphene phonons are also not observed in Raman scattering from monolayer graphene on Ru(0001).[52] The strong film/metal interaction in this system, however, creates a band gap of several eV,[29, 30] sufficiently large to disrupt the electronic transitions involved in the Raman process. In contrast, the perturbation of the electronic bands that Ir produces in the R0 graphene π bands is relatively small (Fig. 3). The major band-structure difference between as-grown R0 and R30 graphene is the transfer of weight to the Ir states from the graphene Fermi surface in the R0 case, which becomes completely suppressed as a result of hybridization. These Ir states occur in a part of momentum space with a very large density of states owing to a van Hove singularity. In this situation, electronic excitations of the Ir states by the graphene phonons will be highly favored, limiting the phonon lifetime and thus destroying the conditions needed to observe Raman scattering in the R0 phase.[51]

3.7 Work functions of R0 and R30 graphene

We further characterize the effect of in-plane orientation by determining the work functions of the R0 and R30 rotational variants from a series of LEEM images as a function of incident electron energy. All electrons are reflected from the surface at low energies. When the energy becomes larger than the work function, electrons begin to be injected into the surface.[53] Figure 7 shows the energy-dependent reflectivities from regions of R0 or R30 graphene. Electrons are injected into R30 graphene at a slightly lower energy than R0 graphene. We measure work function changes using the criteria of the energy that gives 90% reflection.[54] In this manner, we find a small difference between the work functions of R30- and R0-covered Ir(111), with R30 being about ~0.06 eV lower than R0.[55] Using literature values for the work function of clean Ir(111), $\Phi_{Ir}$ = 5.7- 5.76 eV,[56, 57] we determine the work function of the LEEM electron source to be 3.59 – 3.65 eV. Then $\Phi_{R0}$ = 4.86 – 4.92 eV and $\Phi_{R30}$ = 4.80 – 4.86 eV. Thus, the work function of graphene-covered Ir(111) lies between the values of graphite (4.6 eV[58]) and clean Ir. Khomyakov et al.[14] modeled the work function of many metals, but not Ir, covered by monolayer graphene. We next discuss how our findings support R0 graphene being chemisorbed and R30 graphene being physisorbed.

## 4. Discussion and conclusions

Two different in-plane orientations of graphene on the same Ir(111) substrate have significant differences in their electronic bands and Raman scattering. For R30 graphene either as-grown or after intentional K doping, ARPES reveals that the Ir substrate does not significantly perturb the shape of the π bands. As-grown R30 graphene is, however, p-doped by electron transfer to the Ir, more so than the R0 variant. Because of this charge transfer, the work function of R30-covered Ir is lower than clean Ir but larger than graphite (Fig. 7). The Raman-active phonons of R30 graphene are observable while it is still bonded to the Ir. The phonon frequencies suggest some residual compressive strain in the film. All these observations are consistent with R30 graphene being primarily physisorbed on Ir.

In contrast, as-grown R0 graphene exhibits a band gap between the occupied and unoccupied Dirac cones. By varying the photon energy in ARPES and doping with K, we show that the band gap results from hybridization with an Ir state near the Fermi level. This hybridization gives metallic character to the graphene near $E_F$. Since this band gap is removed by doping with adsorbed (non-intercalated) K, the band gap near $E_F$ does not result from any symmetry breaking of the graphene lattice by the substrate.[10] We do find evidence [Fig. 3(c-e, left)] for a very small (~100 meV) band gap is doped R0 graphene, likely from sublattice symmetry breaking. We do not observe the Raman-active phonons in as-grown R0 graphene, suggesting that the hybridization of the π bands near $E_F$ is sufficient to quench the resonant conditions needed to observe the graphene phonons. All these observations show that R0 graphene is primarily chemisorbed on Ir.

The difference in interaction strengths, physisorbed R30 versus chemisorbed R0, helps explain the work-function measurements. From the ARPES results we might expect R0 graphene to have lower work function, contrary to observation (Fig. 7) — the extrapolated R0 π bands cross each other about 75 meV lower (Fig. 3b) than the extrapolated R30 bands (Fig. 5a). That is, the R0 π bands are less *p*-doped than the R30 bands, which for a simple physisorbed system[14] would give R0 the smaller work function.[14] In chemisorbed systems, however, electron transfer from the graphene sigma bonds to the metal can dominate over transfer from the π bonds.[30, 59]

Pletikosić et al. proposed that the R0 graphene bands on Ir(111) were only weakly perturbed because there is a band gap in Ir states near $E_F$ and the graphene K point.[44] If this were true it would be surprising because no such gap occurs near the Dirac cones in the R30 phase, and yet it is the R0 phase that is more hybridized to the metal states. As we have shown, in fact this Ir gap is pinched closed at $E_F$ near the graphene cones, resulting in the complete removal of the graphene Fermi surface in the R0 variant [Fig. 2b-d and Fig. 3(a, right)]. The flat topology of the Ir bands near the Ir saddle point means an anti-crossing between graphene and Ir states can completely remove the graphene Fermi surface. With good reason, then, we speculate that this removal of the Fermi surface is responsible for the relative stability of the R0 variant.

The R30 variant also hybridizes with the Ir states, but in this case the band crossings are between bands of similar slopes, and occur where the graphene Fermi surface has a relatively large extent in momentum space. In these conditions, any anti-crossing between graphene and Iridium bands can remove only a limited part of the Fermi surface.

Such an electronic stabilization of the R0 variant due to hybridization is enhanced over the R30 variant, we speculate, for two additional reasons. First, because of stronger moiré effects, which we explain as follows. Regardless of the relative lattice constants of graphene and Ir, if main Dirac cone lies at a saddle point, then also the satellite cones, must by symmetry exist at equivalent Ir band saddle points and therefore the Fermi surface is also removed for the satellite Dirac cones. (For the same reason the size of the mini-BZ in Fig. 2b will always be such that its K points lie exactly at the Ir's K point). This is a driving mechanism for the C atoms to move in such a way both by buckling and by lattice strain to enhance the strength of the moiré superlattice and to align the Dirac cones with the Ir saddle point. We note that in contrast the superlattice potential is very weak in the R30 variant, as reflected in the lack of satellite features in ARPES measurements and very weak satellite diffraction in LEED.[19]

Second, the Ir bands' saddle point is special because it creates a Van Hove Singularity (VHS) in the Ir electronic density of states. This not only provides an additional explanation of the increased hybridization strength (which is reflected in the wider anti-crossing gap observed for the R0 phase) but also can help stabilize the buckling of the atoms in the moiré superlattice. Such a VHS leads to a strong singularity in the momentum-dependent dielectric function of the Ir surface. Therefore, the electron-electron interactions that might prevent the C atoms from buckling are weakened because of the screening of the Coulomb interactions near the Ir bands' saddle points. The Ir $5d$ bands play two roles in this explanation, first to form the chemical bond through hybridization, and second to provide screening to support the lower symmetry of the incommensurate superlattice.

Ir(111)[60] and Pd(111)[18] are intermediate between strongly and weakly interacting metal substrates for graphene. Their interaction is not strong enough to select a single in-plane orientation, as monolayer graphene on Ni(111)[15] and Ru(0001).[16] But the interaction is strong enough to cause some properties to have an orientation dependence, unlike Pt(111).[20] The strength of the graphene/metal binding in systems that form moirés affects other properties.[13] Consider the progressively weaker binding going from monolayer graphene on Ru(0001) to R0 and then R30 on Ir(111). On Ru(0001), even the high-order superstructure spots from the moiré are intense in LEED[13], the moiré cell has large corrugations in apparent height in scanning tunneling microscopy (STM) images (about 1 Å)[13] and graphene has a large band gap.[10] The chemisorption gives sufficient electron transfer from the graphene to the metal to lower the work function below that of graphite.[59] For R0 graphene on Ir(111), there are still pronounced high-order superstructure spots in LEED and K-point replicas occur in ARPES (Fig. 2). But the STM corrugations are much smaller (about 0.3 Å),[19] as is the band gap (section 3.2) and the work function lies between graphite and Ir. For the R30 variant on Ir, high-order superstructure spots are not found in LEED,[19] there are no K-point replicas in ARPES (Fig. 2) and the STM corrugation is about ten-times yet smaller (about 0.04 Å).[19] Clearly the strength of the band structure replicas and the superstructure diffraction in LEED correlates with the strength of the film/substrate interaction, as measured by the size of the band gap in ARPES or the ability to observe the Raman-active phonons. Understanding these detailed dependencies on film/substrate interaction on the metal and its relative orientation with the graphene should help in the development of metal contacts to graphene devices.

**Acknowledgments**

The authors thank Joshua Whaley for programming the stage of the Raman system. Work at Sandia was supported by the Office of Basic Energy Sciences, Division of Materials Sciences and Engineering of the U.S. DOE under Contract No. DE-AC04-94AL85000. The Advanced Light Source is supported by the Director, Office of Science, Office of Basic Energy Sciences, of the U.S. Department of Energy under Contract No. DE-AC02-05CH11231. L. M. acknowledges support by the Swiss National Science Foundation through project PBELP2-125484.

**Figures**

Fig. 1. (Color online) Brillouin zones of Ir(111) surface and a) R0 and b) R30 rotational variants of graphene. Reciprocal vectors, $a_1^{Ir}$ and $a_2^{Ir}$, of the Ir(111) surface are shown in blue. Reciprocal vectors of graphene honeycomb lattice, $b_1^{R0}$, $b_2^{R0}$, $b_1^{R30}$ and $b_2^{R30}$, are displayed in red. Reciprocal vectors of the superstructures, $c_1^{moiré}$ and $c_2^{moiré}$, are shown in black. The observed (R0) and possible (R30) band-structure replicas are depicted around one graphene K point with black circles.

Fig. 2. (Color online) a). Map of states at a binding energy -330 meV below the Fermi level for an as-grown film containing a mixture of R0 and R30 graphene on Ir(111). Drawn lines show Brillouin zones of Ir(111) surface (blue), R0 graphene (solid red) and R30 graphene (dashed red). Hexagon around $K_{R0}$ point at upper right is the system's mini-Brillouin zone. b-d). Magnified views around $K_{R0}$ at binding energies of -150 meV, -75 meV and at the Fermi level, respectively, showing strong hybridization of the graphene and Ir states near the Fermi level. Photon energy = 95 eV.

Fig. 3. (Color online) Energy bands around a K point of R0 graphene on Ir(111). Left panels). Energy vs. momentum along $\Gamma K_{R0}$ direction. Center panels). Energy vs. orthogonal momentum

direction. Right panels). Map of states at the Fermi level. Black dot marks $K_{R0}$. a). As-grown film using 95 eV photons. b). As-grown film using 130 eV photons. c-e). After light, medium and heavy dosing with potassium, respectively, using 130 eV photons.

Fig. 4. (Color online) Mini-gap spectrum of R0 graphene. a). Momentum distribution slice around the mini-BZ at -770 meV binding energy, where the faint circular contours of the satellite (replica) cones overlap at the mini-BZ boundary, the white hexagon. b-bottom). ARPES spectrum along mini-BZ boundary. Ir bands are marked. b-top). Mini-gap size vs. momentum. Black line is present study and red circles are results of Pletikosić et al.[33] Photon energy = 130 eV.

Fig. 5. (Color online) Left panels). Energy vs. momentum along $\Gamma K_{R30}$ direction. Center panels). Energy vs. orthogonal momentum direction. Right panels). Map of states at the Fermi level. a). As-grown film. b). After heavy dosing with K. Ir bands are marked with dashed lines. Photon energy = 130 eV.

Fig. 6. a) Composite of two LEEM images of graphene-covered Ir(111). Field-of-view is 34 μm × 53 μm. The dark, medium and bright contrasts are bare Ir, R0 graphene and R30 graphene, respectively. The bright horizontal and vertical stripes are scratches used to find the same region in optical microscopy. b). Raw Raman spectra collected from evenly spaced intervals along the red arrow in a). Spectra are offset for clarity, with the bottom and top spectra coming from the start and end of the arrow, respectively. c). Background-subtracted Raman spectra of R0 and R30 graphene. Spectra are offset for clarity. d). G' peak of the R30 variant fitted by a Lorentzian function.

Fig. 7. Electron reflectivity vs. electron energy for Ir(111) either clean or covered by R0 or R30 graphene. The work function is measured at the energy where the reflectivity decreases to 90% of total reflection. Insert: Boxes in LEEM image show regions where R0 and R30 electron reflectivities were measured. Boundary between R0 and R30 graphene runs roughly from lower left to upper right. Smooth vertical lines are Ir steps and step bunches. Field-of-view is 9 μm × 9 μm.

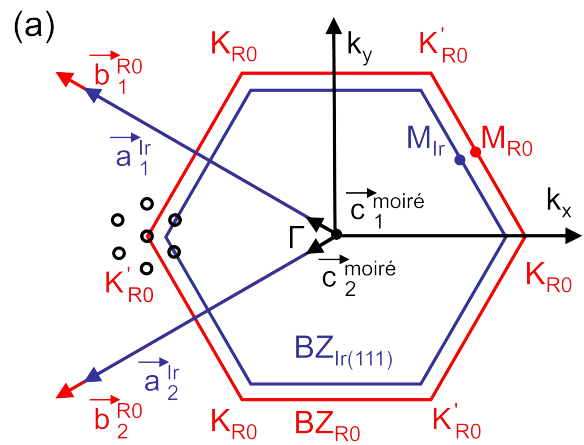
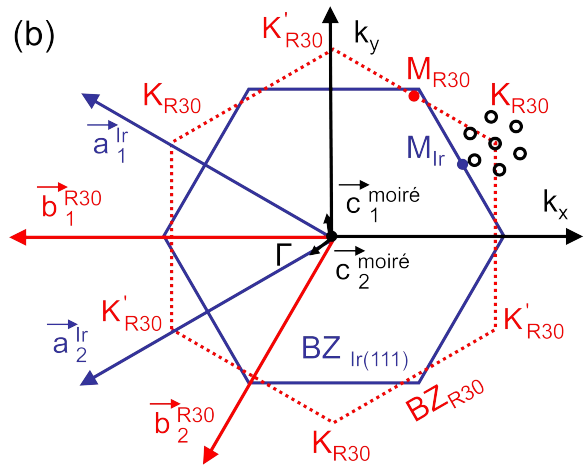

Figure 1

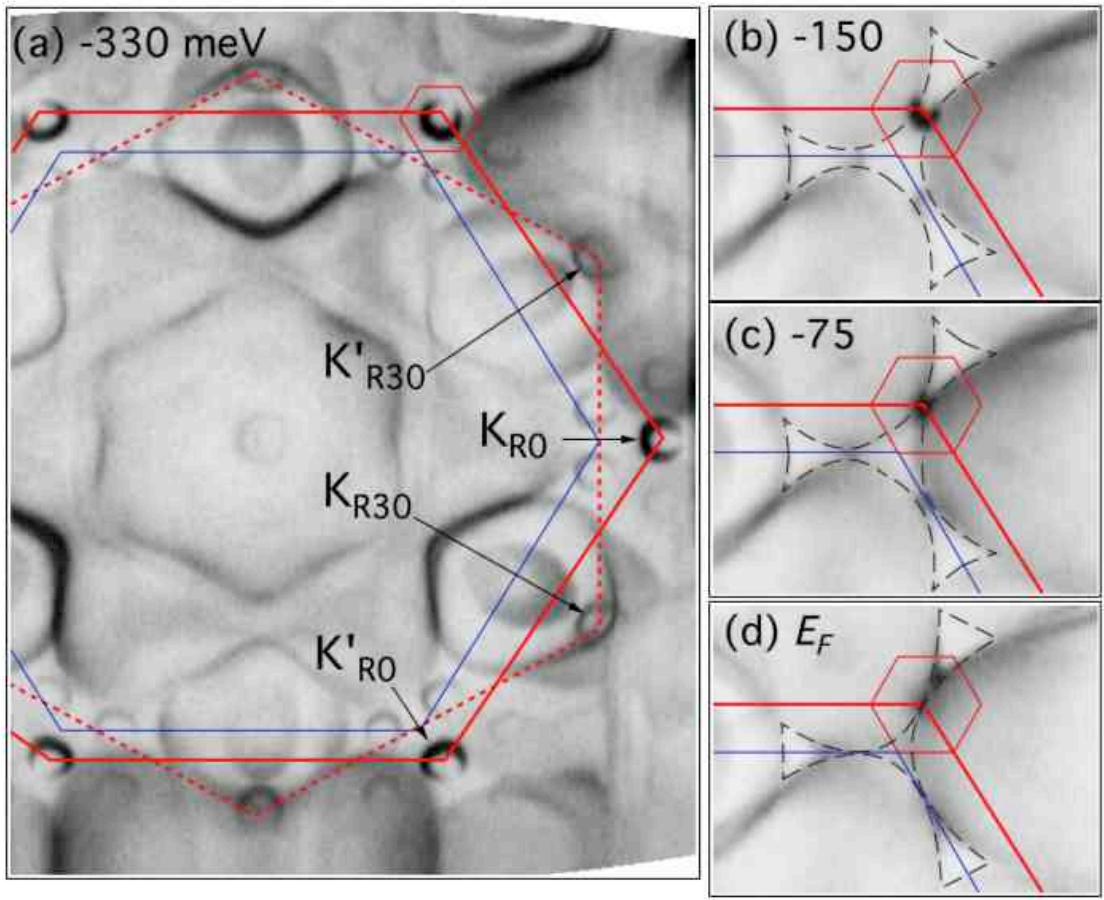

Figure 2

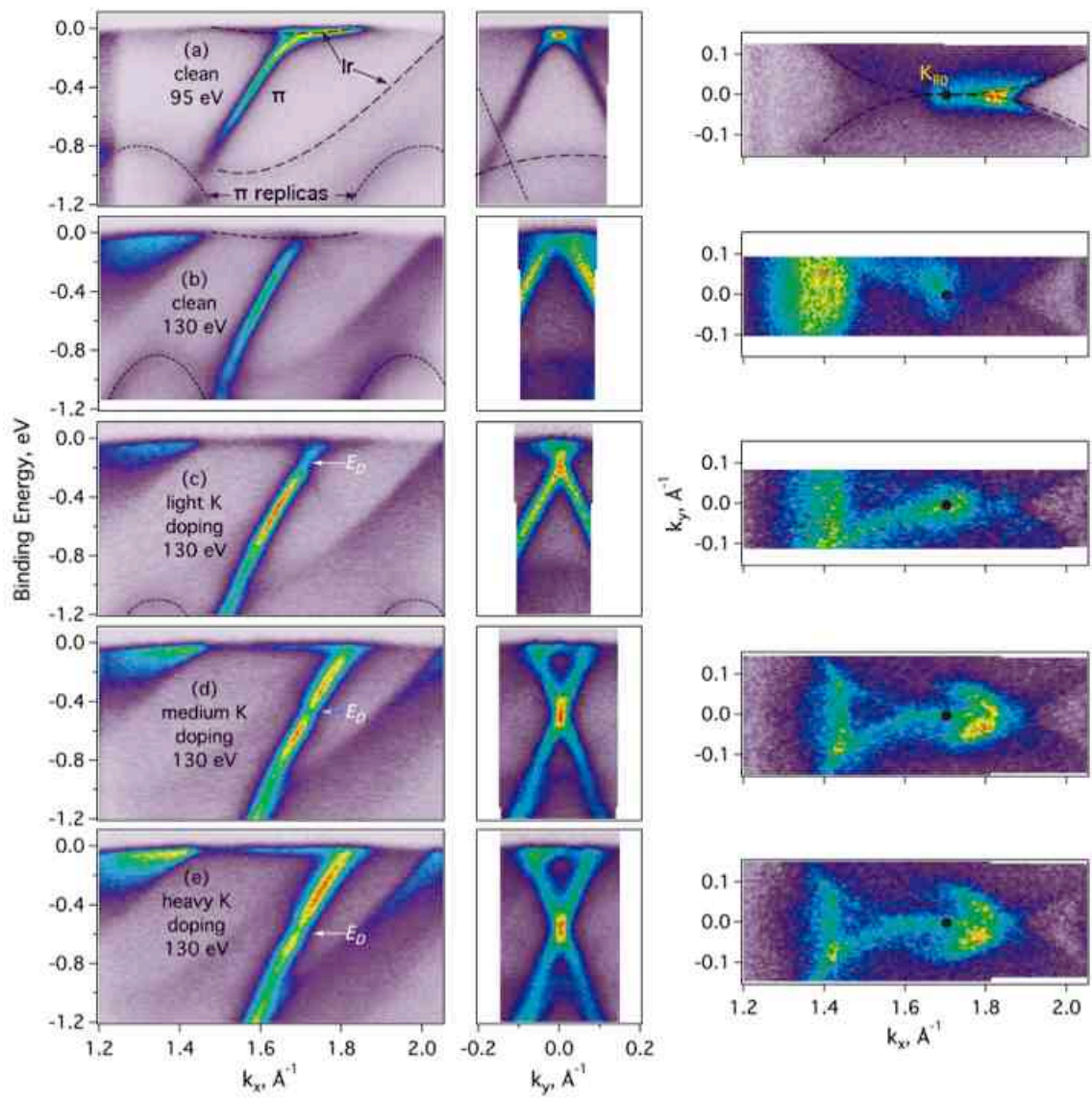

Figure 3

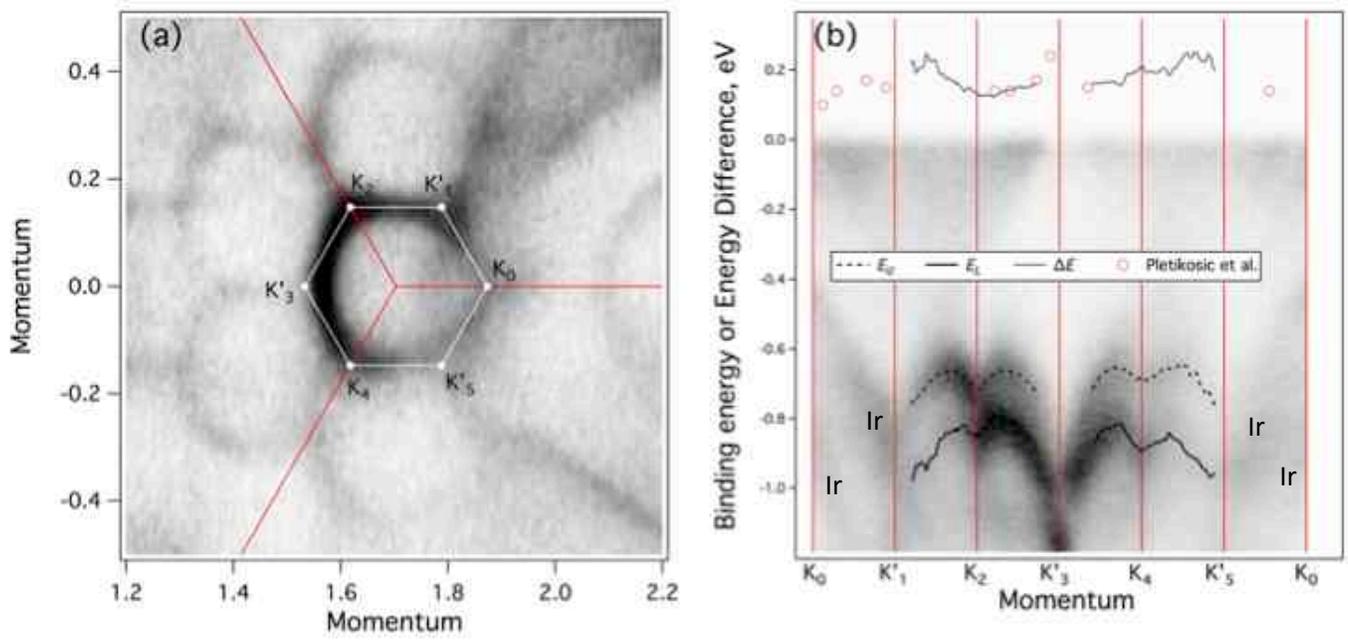

Figure 4

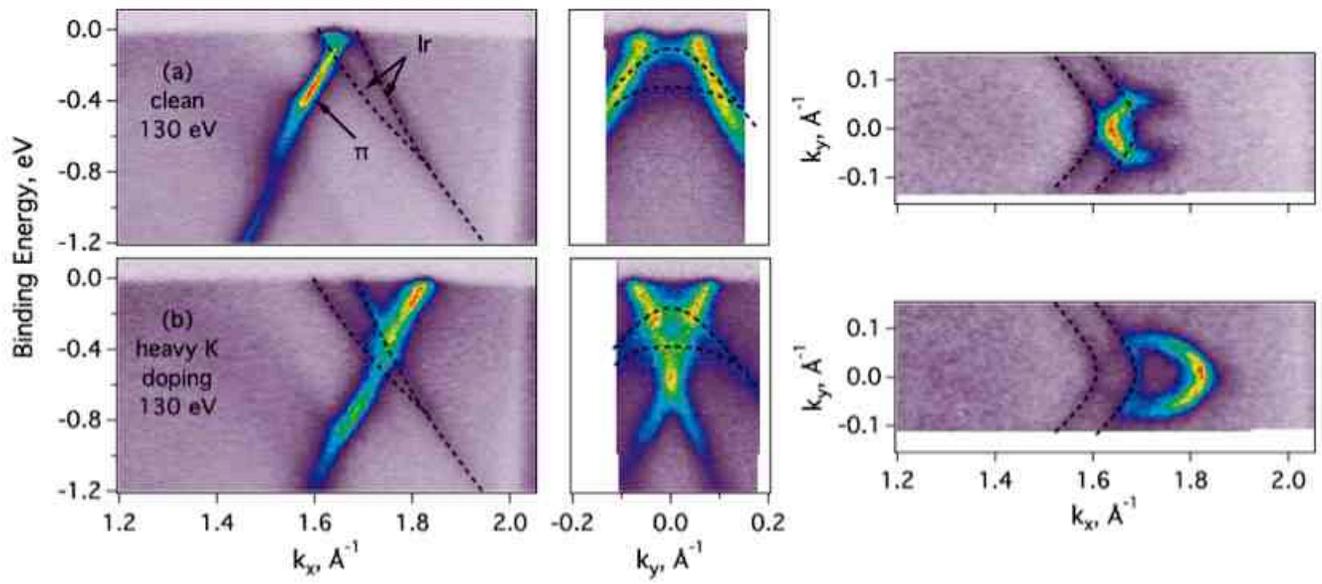

Figure 5

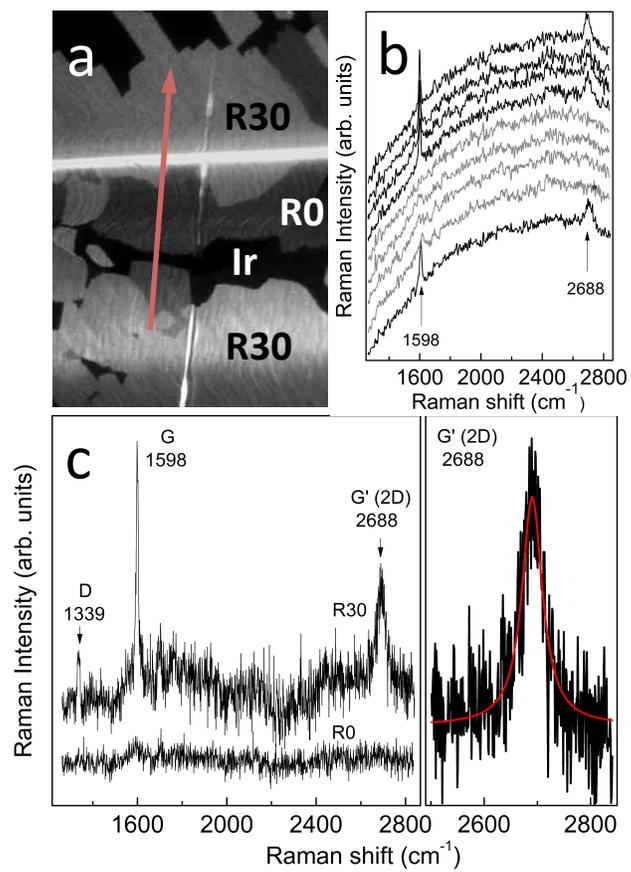

Figure 6

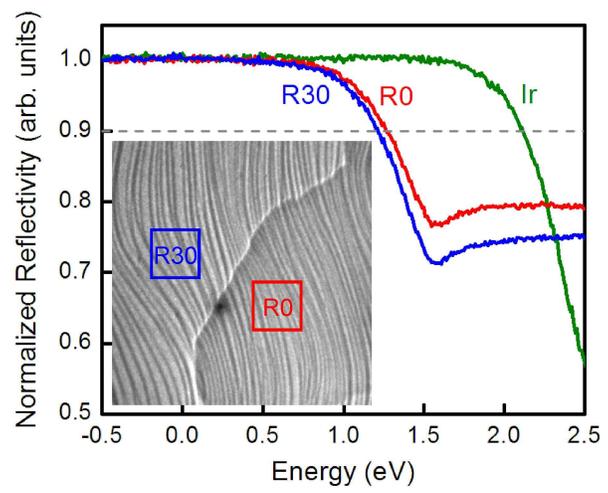